\documentclass[fleqn,usenatbib]{mnras}

\usepackage{newtxtext,newtxmath}

\usepackage[T1]{fontenc}

\DeclareRobustCommand{\VAN}[3]{#2}
\let\VANthebibliography\thebibliography
\def\thebibliography{\DeclareRobustCommand{\VAN}[3]{##3}\VANthebibliography}

\usepackage{graphicx}	
\usepackage{amsmath}	
\usepackage{soul}
\usepackage{subcaption}
\usepackage{xcolor}

\newcommand{\thd}[3]{\left(\frac{\partial #1}{\partial #2}\right)_{#3}} 
\newcommand{\td}[2]{\frac{d #1}{d #2}} 

\title[PNS Convection and NDW $\nu$p-Processing]{Proto-Neutron Star Convection and the Neutrino-Driven Wind: Implications for the $\nu$p-Process}

\author[B. Nevins and L. Roberts]{
Brian Nevins,$^{1,2,3}$\thanks{E-mail: nevinsb1@msu.edu}
Luke~F. Roberts$^{4}$
\\
$^{1}$Department of Physics and Astronomy, Michigan State University, East Lansing,
MI 48824, USA\\
$^{2}$Facility for Rare Isotope Beams, Michigan State University, East Lansing, MI
48824, USA\\
$^{3}$Joint Institute for Nuclear Astrophysics – Center for the Evolution of the Elements (JINA-CEE), USA\\
$^{4}$Computer, Computational, and Statistical Sciences Division, Los Alamos National Laboratory, Los Alamos, NM, 87545, USA}

\date{Accepted XXX. Received YYY; in original form ZZZ}

\pubyear{2022}

\begin{document}
\label{firstpage}
\pagerange{\pageref{firstpage}--\pageref{lastpage}}
\maketitle

\begin{abstract}
Recent studies of the neutrino-driven wind from proto-neutron stars have indicated that the wind is likely proton-rich for much of its lifetime, and the high flux of neutrinos can induce $\nu$p-process nucleosynthesis allowing for the formation of heavy elements. It has also been shown that gravito-acoustic waves, generated by convection within the proto-neutron star, can significantly alter the dynamics and nucleosynthesis in the wind. Therefore, we present a study of the effects of convection-driven waves on the nucleosynthesis in proton-rich neutrino-driven winds, focusing on the $\nu$p-process. We find that wave effects can strongly impact $\nu$p-process nucleosynthesis even at wave luminosities a factor of $10^{-5}$ smaller than the total neutrino luminosity. The momentum flux of the waves accelerates the wind, reducing the net neutrino heating and the persistent neutron abundance created by p($\bar{\nu}_e,e^+$), which impedes $\nu$p-process nucleosynthesis. However, this effect is generally counteracted by the effects of waves on seed nucleus formation, as the acceleration of the wind and the heating that occurs as these waves shock both favor a more $\alpha$-rich environment with very little heavy seed-nucleus formation. Overall, higher wave luminosities correlate (albeit non-monotonically) with heavier element $\nu$p-processing, up to $A\approx 200$ in some cases. At very high wave luminosities ($\gtrsim 10^{-3}L_\nu$), early shock heating by the waves disrupts $\alpha$ recombination, and drives a suppressed, fast-outflow r-process proceeding up to $A\approx 200$. This occurs despite an assumed neutrino spectrum that predicts a proton-rich wind with equilibrium $Y_e=0.6$.
\end{abstract}




\section{Introduction}

As a newly formed proto-neutron star (PNS) cools, the immense neutrino flux liberates some material from its surface as a neutrino-driven wind \citep[NDW, see][]{Duncan_1986}. Material ejected from the surface of the PNS initially consists of free protons and neutrons, with the proton fraction ($Y_p = Y_e$ where $Y_e$ is the electron fraction) of the NDW set by the spectra of the electron neutrinos and antineutrinos driving it. As material in the wind expands and cools, the neutrons and protons combine to form alpha particles, leaving behind some excess protons or neutrons, depending on whether $Y_e > 0.5$. These alpha particles then combine through a triple-$\alpha$ process followed by $\alpha$ capture reactions to form heavier ``seed'' nuclei. The free nucleons can then capture onto the seed nuclei and produce either an (weak) r- or $\nu$p-process \citep[see][for a recent review]{Arcones_2012}.

The dynamics of the wind play an important role in determining how how high in atomic mass these nucleosynthetic processes can proceed. High entropies and fast expansion timescales during the seed-formation phase inhibit the 3- and 4-body reactions that form the initial heavy nuclei which can then undergo capture reactions \citep[e.g.][]{Hoffman_1996,Hoffman_1997}. A decreased seed abundance allows for more free nucleons to capture onto each seed, and nucleosynthesis can proceed to higher mass numbers. A number of physical phenomena affecting the dynamics of the wind have been explored in previous studies. General relativistic effects can increase the entropy of the wind significantly \citep{Cardall_1997,Otsuki_2000,Thompson_2001}. PNS rotation and magnetic fields can decrease the dynamical timescale \citep{Metzger_2007,Vlasov_2014,Thompson_2018,Desai_2022}. Another important factor, which we focus on here, is the presence of propagating waves in the NDW \citep[e.g.][]{Suzuki_2005,Dessart_2006,Metzger_2007,Gossan_2020}. In \citet{Nevins_2023}, henceforth Paper I, we explored the influence of gravito-acoustic waves driven by convection in the PNS \citep{Gossan_2020} on a possible r-process in a mildly neutron-rich NDW, and found that even a small wave luminosity could drive a strong third-peak r-process in winds that otherwise could not move beyond the first peak. In this paper, we turn our attention to the $\nu$p-process in a proton-rich wind. 

While there remains substantial uncertainty in the neutrino spectrum of cooling PNSs, several recent studies have indicated that the wind becomes increasingly proton-rich at late times \citep[e.g.][]{Martinez_Pinedo_2014,Fischer_2020,Pascal_2022}. Convection in the PNS, which \citet{Nagakura_2021} recently showed to be significant across a broad range of progenitors for the first few seconds of PNS lifetime, also seems to lead to more proton-rich NDWs \citep{Pascal_2022}. This may be counteracted, however, by the mechanical effects of convection-driven waves accelerating the NDW, reducing neutrino captures and leading to a more neutron-rich wind (Paper I). Nonetheless, the growing consensus over the last decade has been that the NDW is likely to be either slightly neutron-rich or proton-rich at earlier times, becoming increasingly proton-rich at late times.

While a proton-rich NDW may be unable to produce heavy nuclei via the r-process, the $\nu$p-process may be active instead. The $\nu$p-process was first proposed by \citet{Frohlich_2006} as a primary nucleosynthetic process that is active in neutrino-rich environments like the NDW. The $\nu$p-process, similar to the rp-process, forms heavy nuclei by proton captures onto seed nuclei followed by beta decays toward stability. Unlike the r-process, which can proceed to the dripline relatively uninhibited regardless of atomic number, the rp-process proceeds closer to stability. This results in longer $\beta$-decay lifetimes for certain waiting-point nuclei, and when charged-current neutrino reactions are neglected it is impossible for the classical rp-process to form nuclei heavier than $^{64}$Ge in the NDW due to its long $\beta$ decay lifetime \citep{Pruet_2005}.  Waiting-point nuclei like $^{64}$Ge can be bypassed, however, by capturing free neutrons created by charged-current antineutrino interactions on free protons. This modified rp-process, called the $\nu$p-process, potentially allows for formation of heavy elements up to $A=100$ and beyond with even a small flux of free neutrons. Since the primary nucleosynthesis flows will be on the proton-rich side of stability, this process could be the source of p-nuclei such as $^{92,94}$Mo and $^{96,98}$Ru. The source of these nuclei and their relative abundances have been problematic for some time \citep[e.g.][]{Arnould_2003}, but a $\nu$p-process in the NDW has been shown to be a viable source for at least some of these nuclides \citep{Wanajo_2006b,Bliss_2018b}.

A number of parameter studies have explored how the hydrodynamic conditions of the wind affect a possible $\nu$p-process \citep[e.g.][]{Pruet_2006,Wanajo_2006b,Wanajo_2011,Arcones_2012b,Eichler_2018}. As described above, the entropy and dynamical timescale of the wind play key roles in determining the ratio of free nucleons to seeds, which places a natural limit on how heavy of nuclei can be formed \citep{Wanajo_2006b,Wanajo_2011}. The electron fraction also affects this ratio, both by dictating how many free nucleons are left after $\alpha$ recombination (assuming a complete $\alpha$ recombination), and by determining the number of free protons which can undergo antineutrino capture and transform into neutrons \citep{Wanajo_2011,Eichler_2018}. In Paper I, we found that the presence of gravito-acoustic waves can substantially increase the entropy of the wind once the waves begin to shock, possibly allowing a $\nu$p-process to proceed to higher mass numbers as seed formation is inhibited. However, the additional heating and momentum flux provided by the waves had the additional effects of reducing the dynamical timescale of the wind and reducing the electron fraction, driving the wind towards neutron-richness. The reduced electron fraction will likely inhibit a strong $\nu$p-process by reducing the potential neutron flux. The effect of a faster outflow is not immediately clear. While it will reduce the efficiency of seed formation (see Paper I), it will also reduce the amount of possible antineutrino captures in the wind, and thus the available neutron flux. A third key factor is the wind termination shock, when the NDW impacts the slower-moving ejecta behind the main supernova shock. The $\nu$p-process is most efficient at temperatures between 1.5 and 3 GK, and a termination shock within this temperature range will prolong efficient $\nu$p-processing. 

Our aim in this paper is to explore the effect of gravito-acoustic waves on heavy element nucleosynthesis in proton-rich winds. We anticipate this will primarily affect nucleosynthesis via the entropy and electron fraction, but we will also briefly explore the role of the wind termination shock in the presence of these waves. We are mainly interested in two aspects of heavy nucleosynthesis: how the active nucleosynthetic processes are affected by the presence of these waves; and what the heaviest elements are that can be formed in a proton-rich wind affected by gravito-acoustic waves. Understanding the impact of waves on the $\nu$p-process has the potential to impact our understanding of both the origin of the heavy elements generally, and the origin of the light p-nuclides $^{92,94}$Mo and $^{96,98}$Ru specifically. Given the significant impact these waves can have in a marginally neutron-rich wind, up to driving a full solar-like r-process, we present here a similar parameter study considering how these waves affect a proton-rich wind and the strength of the $\nu$p-process there. In section \ref{sec:theory} we describe specifically how we expect the presence of these waves to affect the $\nu$p-process. Section \ref{sec:model} describes the modeling method we use to investigate this, and section \ref{sec:results} shows the results of our simulations. We find that the presence of gravito-acoustic waves in the NDW has a substantial impact on $\nu$p-process nucleosynthesis and can allow much heavier elements to be formed than would otherwise be possible.

\section{Wave Effects and the $\nu$p-process}\label{sec:theory}

In Paper I, we provide a detailed description of the mechanism by which gravito-acoustic waves are generated and subsequently alter the dynamics of the wind. To summarize, convective regions within the PNS generate gravity waves, which tunnel through an evanescent region near the surface of the PNS and emerge as acoustic waves in the NDW. Eventually, the waves will steepen into shocks, which will efficiently deposit heat into the wind according to
\begin{equation}
    \dot{q}_w = \frac{L_w}{4\pi r^2 \rho l_d} = \frac{c_s}{\rho l_d}\omega S.
\end{equation}
Here $r$ represents the radius of a fluid element, $v$ the velocity, $c_s$ the local sound speed, and $\rho$ the density. $S$ is the wave action \citep{Jacques_1977}, and $l_d$ is the dissipation length of the waves, given by
\begin{equation}
    l_d = \frac{\pi\gamma^2}{\gamma+1}\left(\frac{c_s^2\epsilon}{3\omega^3 S}\right)^{1/2}
\end{equation}
where $\gamma$ is the adiabatic index and $\epsilon$ the energy density of the wind. This determines how rapidly heat is deposited into the wind once shock heating begins. In terms of the wave luminosity $L_w$, the dissipation length will scale as
\begin{equation}
    l_d \propto L_w^{-1/2}\omega^{-1}.
\end{equation}
The radial distance $R_s$ at which these shocks will form depends on both the wave luminosity $L_w$ and the frequency $\omega$ of the waves, and can have an important impact on nucleosynthesis. Following \citet{Mihalas_1984}, the exact value of $R_s$ is determined by
\begin{equation}\label{eq:rshock}
    \int_0^{R_s} \sqrt{\frac{\omega S}{\rho}}dr = \frac{2\pi c_s^2}{\omega(\gamma+1)}.
\end{equation} 
Approximating density and pressure with power law expressions derived from our models in Paper I, we estimate the radius of shock formation will scale as
\begin{equation}
    R_s \propto L_w^{-2/11}\omega^{-4/11}.
\end{equation}

Seed formation typically begins when the wind cools to approximately $T$=0.5 MeV; if shock formation occurs before or near this point it can drastically reduce the number of seeds that form. \citet{Dessart_2006} and \citet{Gossan_2020} predict frequencies of $10^2 - 10^4$ rad s$^{-1}$, and in Paper I we found that wave frequencies of $\sim 10^4$ rad s$^{-1}$ resulted in shock formation early enough to strongly inhibit seed formation and drive a robust r-process for fiducial PNS parameters. We refer the reader to Paper I for a more thorough description of these effects. We estimate, as in Paper I, that the luminosity carried in these waves will obey the scaling relationship
\begin{equation}
    L_w \sim M_{\text{con}}\mathcal{T}L_{\nu},
\end{equation}
with $M_{\text{con}}$ representing the convective Mach number in the PNS, $\mathcal{T}$ representing the transmission coefficient through the evanescent region, and $L_\nu$ representing the total neutrino luminosity of the PNS. Based on results from \citet{Dessart_2006} and \citet{Gossan_2020}, we estimate a wave luminosity in the wind of order $10^{-5}$ -- $10^{-2} L_\nu$. 

Acoustic waves propagating through the NDW will accelerate it, and when the waves shock they will also raise the entropy of the material. Shorter dynamical timescales (i.e. faster outflows) and higher entropies favor an alpha-rich freezeout and a smaller number of seed nuclei \citep{Hoffman_1997}, which will allow the $\nu$p-process to form heavier nuclei. Both the proton-to-seed ratio and the efficiency of neutron production depend also on the free proton abundance, which is impacted by the  waves. Because the wind is being accelerated by the waves, less neutrino heating is needed to unbind the wind material. This means that the material may not reach the equilibrium electron fraction ($Y_{e,\text{eq}}$) predicted by the PNS neutrino spectrum \citep[see][]{Qian_1996}, but will stay closer to neutron-richness (Paper I). The lowered proton fraction will reduce the proton-to-seed ratio (disfavorably for a strong $\nu$p-process, in contrast to the impact of the acceleration and entropy deposition on seed formation). The wave acceleration may also reduce the number of antineutrino captures that take place, reducing the free neutron flux and again negatively impacting a $\nu$p-process. 

\section{Modeling Method}\label{sec:model}

We calculate the NDW structure using a general relativistic, spherically symmetric, steady state model that includes the impact of gravito-acoustic wave acceleration and heating. As described in Paper I, number conservation, momentum conservation and energy conservation in the time-independent outflow lead to the wind equations for the mass outflow rate $(\dot M)$, radial velocity ($v$), entropy ($s$), electron-fraction ($Y_e$), and the wave action $(S)$ as a function of radius given by  
\begin{eqnarray}\label{eq:critical_form}
\dot M &=& 4 \pi r^2 e^\Lambda W v \rho \nonumber\\
 \td{v}{r} &=& \frac{v}{r}\frac{f_2}{f_1} \nonumber\\
 \td{s}{r} &=& \frac{\xi_s}{r} \nonumber\\
 \td{Y_e}{r} &=& \frac{\xi_{Y_e}}{r} \nonumber\\
 \td{S}{r} &=& -S\left(\frac{2}{r}+\frac{1}{l_d}+\frac{1}{v_g}\td{v_g}{r}\right)
\end{eqnarray}
where
\begin{align}\label{eq:dimensional_quantities}
&f_1 = \left(1 - \frac{v^2}{c_s^2} \right) +\delta f_1 \nonumber\\
&f_2 = - \frac{2}{W^2} + \frac{G M_\text{NS}}{c_s^2 r} \frac{1-\left(\frac{c_s}{c}\right)^2}{e^{2\Lambda} W^2} 
 \nonumber\\&\hspace{20pt} +  \frac{1}{W^2 h \rho c_s^2} \left[\xi_s\thd{P}{s}{\rho,Y_e}+\xi_{Y_e}\thd{P}{Y_e}{\rho,s}\right] + \delta f_2\nonumber\\
&\xi_{s} = \frac{r}{v} \frac{\dot{q}_\textrm{tot}}{e^\Lambda W T} \nonumber\\
&\xi_{Y_e} = \frac{r}{v} \frac{\dot{Y}_e}{e^\Lambda W}.
\end{align}
The wave action $S$ is connected to the wave luminosity $L_w$ via
\begin{equation}\label{eq:wave_lum}
    S = \frac{L_w}{4\pi r^2 c_{s} \omega}.
\end{equation}
The heating rate per baryon $\dot{q}_\textrm{tot} = \dot{q}_\nu + \dot{q}_w$ includes contributions from neutrino heating and wave heating. The electron fraction evolves due to charged-current neutrino capture rate $\dot{Y}_e$. Other quantities are the Lorentz factor $W$, $e^\Lambda=\sqrt{1-\frac{2GM_{NS}}{r c^2}}$, the sound speed $c_s$, the group velocity $v_g$, the density $\rho$, the temperature $T$, the specific enthalpy $h$, the gravitational constant $G$, and the speed of light $c$. The quantities $\delta f_1$ and $\delta f_2$ constitute corrections due to wave stresses and are described in detail in Paper I. The system is closed by the \cite{Timmes_2000} equation of state. We also include a wind termination shock that is parameterized by the radius $r_\text{ts}$ at which the termination shock occurs. Physically, this is the point at which the transonic NDW impacts slower-moving ejecta behind the primary supernova shock. The post-termination shock conditions are then determined by the relativistic Rankine-Hugoniot shock jump conditions, after which we assume a homologous outflow.  The critical form equation has a singularity at the transonic point, which corresponds to the wind solution of the equations (as opposed to a breeze that falls back onto the PNS). We employ a shooting method to find this transonic solution in the steady-state approximation.

From these steady state models, we extract time-dependent histories of temperature, density, and neutrino properties for Lagrangian observers in the flow. Nucleosynthesis is then calculated for these trajectories using the nuclear reaction network SkyNet \citep{Lippuner_2017}. The reaction network calculations include strong, weak, symmetric fission, and spontaneous fission reactions taken from the REACLIB reaction library \citep{Cyburt_2010}, with inverse reactions calculated via detailed balance. In addition, we include neutrino and electron captures on protons and anti-neutrino and positron captures on neutrons using the neutrino properties extracted from the wind models. These rates were not included in Paper I, as they do not have a significant impact on the r-process after $Y_e$ has been set, but they are key to capturing the $\nu$p-process nucleosynthesis. 

The parameter space we explore is identical to that in Paper I, with a proto-neutron star mass $M_\text{NS}/M_\odot \in [1.4, 2.1]$, a total neutrino luminosity in {\it all} flavors of $L_\nu \in$ [$3\times 10^{51}, 1.2\times 10^{53}$] erg s$^{-1}$, and $L_w/L_\nu \in [10^{-5}, 10^{-2}]$. These parameters are varied independently; no correlations between parameters are assumed in this work. We take $M_\text{NS}=1.5 M_\odot$ and $L_\nu=3\times 10^{52}$ as fiducial values for comparison. Electron neutrinos are assumed to have an energy of 12 MeV, and electron antineutrino energies are again calculated such that the equilibrium $Y_e$ \citep{Qian_1996} is equal to a set value, here set to $Y_{e,\text{eq}}=0.6$ unless otherwise stated. We also assume a wind termination shock radius $r_\text{ts}=5\times 10^8$cm and a wave frequency  $\omega=2000$ rad s$^{-1}$ unless otherwise indicated. 

\section{Results}\label{sec:results}
\subsection{Models without Wave Effects}
\begin{figure}
    \centering
    \includegraphics[scale=1]{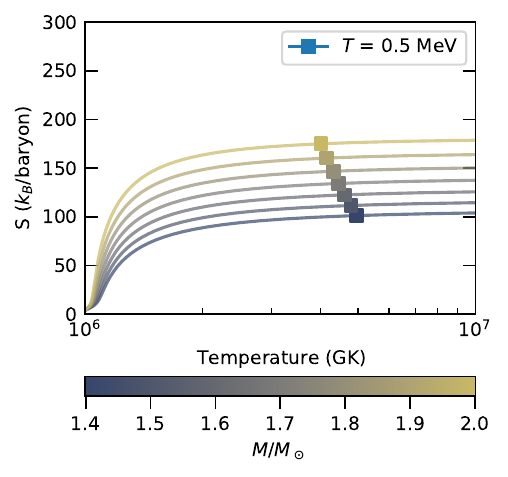}
    \caption{Entropy versus radius for PNSs of varying mass, with the total neutrino luminosity (in all flavors) fixed to a fiducial value of $3\times 10^{52}$ erg s$^{-1}$, and $L_w = 0$.}
    \label{fig:no_heat_entropy}
\end{figure}
\begin{figure}
    \centering
    \includegraphics[scale=1]{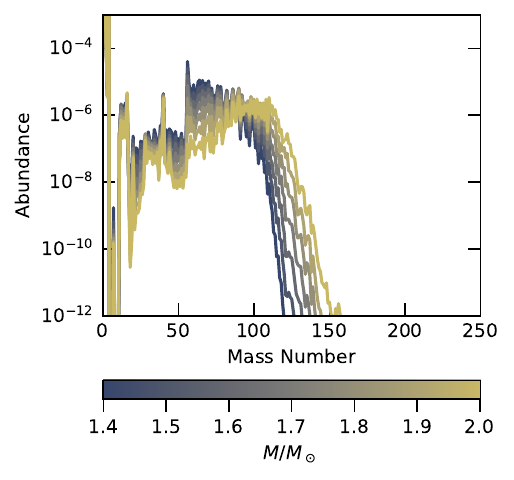}
    \caption{Final abundances for the wind profiles shown in figure \ref{fig:no_heat_entropy}, in the absence of wave effects (i.e. $L_w=0$). We find some limited heavy element formation up to about A=130, with modest PNS mass dependence. Higher PNS masses lead to higher asymptotic entropies in the wind, which are generally favorable for heavier element nucleosynthesis.}
    \label{fig:no_heat_abundance}
\end{figure}
As a baseline for comparison, in figures \ref{fig:no_heat_entropy} and \ref{fig:no_heat_abundance} we show the wind entropy and final abundance distributions for a set of winds absent wave effects. We find, as in Paper I, the expected dependence of asymptotic entropy on mass. Our nucleosynthesis calculations are in general agreement with \citet{Pruet_2006} and \citet{Wanajo_2011}, with heavy element synthesis terminating around $A$=120 for high entropy winds.

\subsection{Models with Wave Effects}
\begin{figure*}
    \centering
    \includegraphics[scale=1]{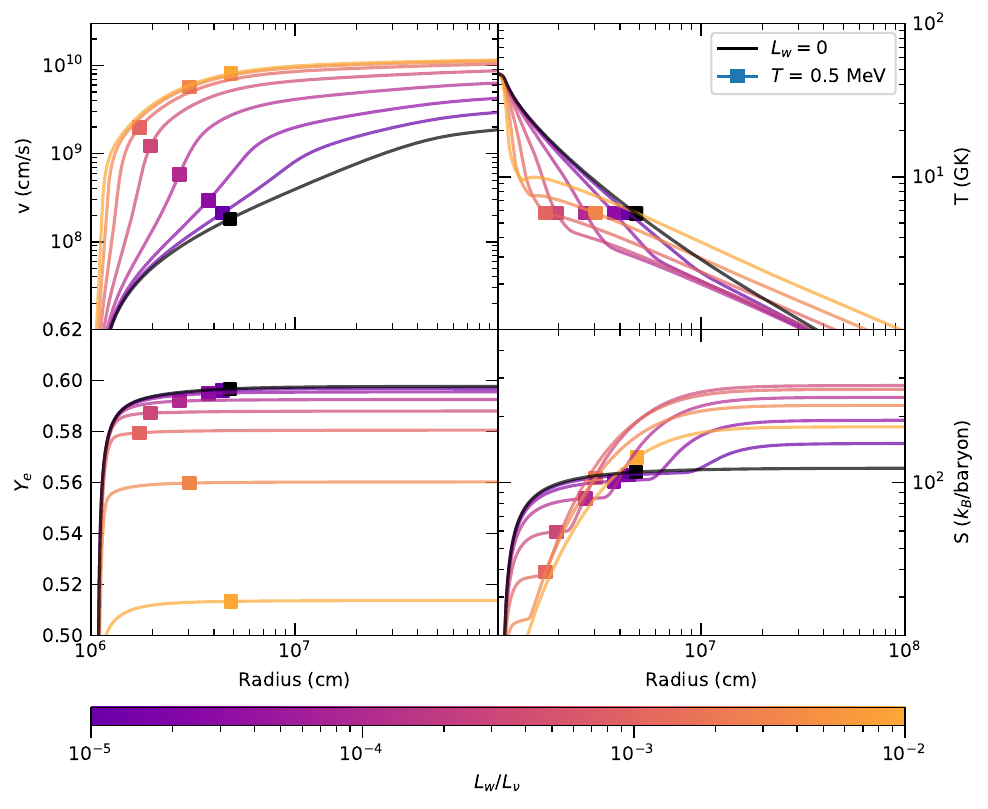}
    \caption{Radial profiles of the velocity, electron fraction, temperature, and entropy in the NDW. Different lines correspond to different $L_w$. Other parameters in the wind models were fixed to $M_\text{NS} = 1.5 \, M_\odot$, $L_\nu = 3 \times 10^{52} \, \textrm{erg s}^{-1}$, and $\omega = 2 \times 10^3 \, \textrm{rad s}^{-1}$. The approximate beginning of seed formation at $T=0.5$~MeV for each model is marked with a square.}
    \label{fig:fid_profiles}
\end{figure*}
\subsubsection{Wind Dynamics}\label{sec:dynamics}
Figure \ref{fig:fid_profiles} shows a set of profiles describing the behavior of the NDW in the presence of wave effects for a fiducial parameter set. Our results are expectedly similar to those in Paper I, with the exception of the increased electron fraction due to the antineutrino energies we assume here. As the wind is heated and accelerated by high wave luminosities, the number of neutrino captures required to unbind the material decreases, preventing the electron fraction from reaching its neutrino capture rate-equilibrium value of $Y_{e,\text{eq}}=0.6$. At the very highest wave luminosities, these winds become almost neutron-rich, despite the assumed neutrino spectrum. This reduction in $Y_e$ will have a substantial impact on the $\nu$p-process in these models, as fewer free protons will be available to produce neutrons via antineutrino captures after seed nucleus formation \citep{Wanajo_2011}. 

Another important feature to note is the onset of shock heating, visible as a sharp increase in the entropy. For higher wave luminosities, this takes place near or before the onset of seed nucleus formation at approximately 6 GK. We also see a clear acceleration of the wind as the wave luminosity is increased, reducing the dynamical timescale over which seed nuclei can form. Results for a more extreme parameter set, with higher mass and $L_\nu$, are qualitatively similar, with one notable difference being higher $Y_e$ at large $L_w$. Increasing the neutrino luminosity can counteract the reduction in neutrino captures caused by high $L_w$, pushing the electron fraction closer to its equilibrium value. A high wave luminosity coupled with a low neutrino luminosity could result in a neutron-rich wind, even though the neutrino spectrum dictates a proton-rich equilibrium $Y_e$.

\subsection{Nucleosynthesis}
\begin{figure*}
\begin{subfigure}{.5\textwidth}
    \includegraphics[scale=1]{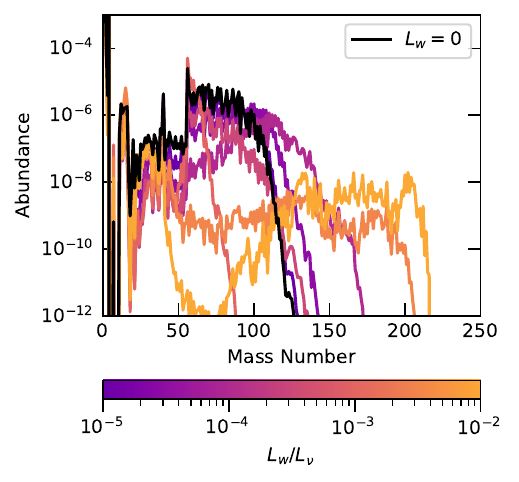}
    \caption{Fiducial parameter set}
    \label{fig:fid_skynet}
\end{subfigure}%
\begin{subfigure}{.5\textwidth}
    \includegraphics[scale=1]{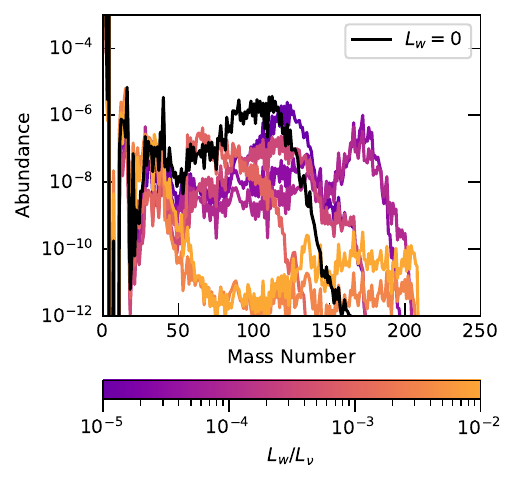}
    \caption{Extreme parameter set}
    \label{fig:ext_skynet}
\end{subfigure}
\caption{Final nucleosynthesis patterns for varied wave luminosities, with other parameters held constant. We assume a wave frequency of $\omega=2\times 10^3$~rad~s$^{-1}$ for both panels. Left: a fiducial parameter set with $M_\text{NS} = 1.5 \, M_\odot$ and $L_\nu = 3 \times 10^{52} \, \textrm{erg s}^{-1}$. Right: a more extreme parameter set with $M_\text{NS} = 1.9 \, M_\odot$ and $L_\nu=6\times 10^{52}$ erg s$^{-1}$. For both parameter sets, we observe two distinct families of abundance patterns. For $L_w/L_\nu\lesssim 2\times 10^{-4}$, the nucleosynthesis is characteristic of a $\nu$p-process, which can proceed to very high mass numbers in the high-entropy environment of a massive, 1.9~$M_\odot$ PNS. At higher wave luminosities, the nucleosynthesis shifts to a suppressed r-process pattern reminiscent of that predicted by \citet{Meyer_2002}.}
\end{figure*}
\begin{figure}
    \centering
    \includegraphics[scale=1]{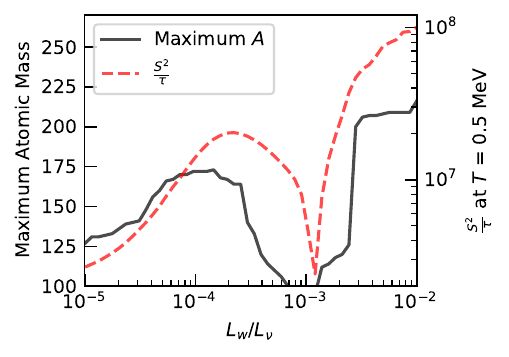}
    \caption{A comparison of the diagnostic quantity $s^2/\tau$, evaluated at $T=0.5$ MeV, with the maximum atomic mass produced with a final abundance greater than $10^{-12}$, for the same PNS parameters ($M_\text{NS}$, $L_\nu$, $\omega$) as in figure \ref{fig:fid_profiles}. We see that the strength of the $\nu$p-process is correlated with this quantity, with the fast-outflow r-process seeming to appear once it crosses a certain threshold value at high $L_w$. }
    \label{fig:s3tau}
\end{figure}
\begin{figure}
    \centering
    \includegraphics[scale=1]{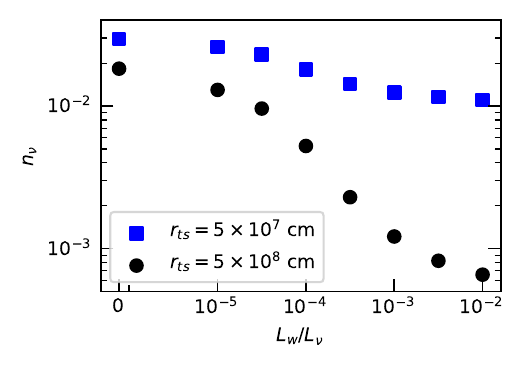}
    \caption{
    Time integrated antineutrino capture rates per baryon ($n_\nu$) from $T$~=~3~GK through the end of the simulation \citep[see][]{Pruet_2006}.  The black dots correspond to the wind profiles shown in figure~\ref{fig:fid_profiles}, which have a wind termination shock imposed at $r_\text{ts} = 5 \times 10^8 \, \textrm{cm}$, while the blue squares show $n_\nu$ for the same set of models but with a wind termination shock imposed at $5 \times 10^7 \, \textrm{cm}$. In both cases, the additional momentum flux and shock heating from the waves reduces the number of neutrino captures as the wave luminosity is increased. An earlier wind termination shock causes the material to remain at smaller radii longer and undergo more neutrino captures.}
    \label{fig:nv}
\end{figure}
\begin{figure*}
\begin{subfigure}{.5\textwidth}
    \centering
    \includegraphics[scale=1]{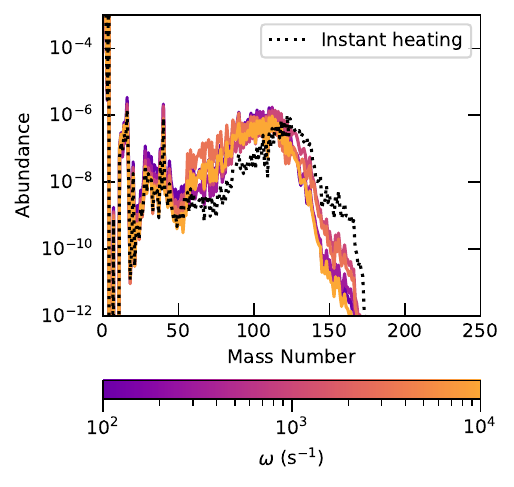}
    \caption{$\nu$p-process ($L_w/L_\nu=1\times 10^{-4}$)}
    \label{fig:freq_vp}
\end{subfigure}%
\begin{subfigure}{.5\textwidth}
    \centering
    \includegraphics[scale=1]{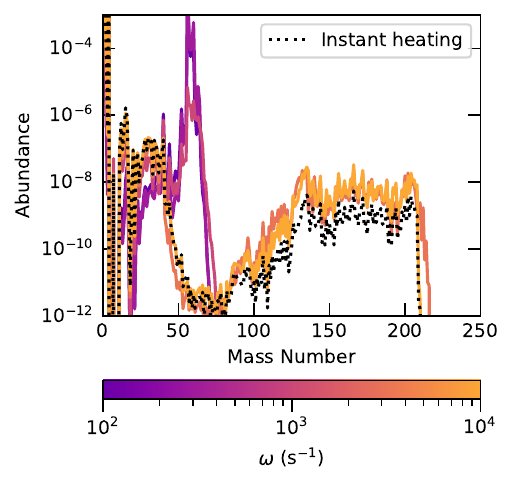}
    \caption{r-process ($L_w/L_\nu=5\times 10^{-3}$)}
    \label{fig:freq_r}
\end{subfigure}
\caption{Final nucleosynthesis results for a fiducial 1.5~$M_\odot$ PNS with $L_\nu=3\times 10^{52}$ erg s$^{-1}$, highlighting the frequency ($\omega$) dependence of the two nucleosynthetic processes. Left: $L_w/L_\nu=1\times 10^{-4}$ to produce a $\nu$p-process. The final abundance pattern depends very weakly on wave frequency. Right: $L_w/L_\nu=5\times 10^{-3}$ to produce a fast-outflow r-process. The fast-outflow r-process requires a minimum wave frequency of approximately $\omega=10^{3}$~s$^{-1}$ to operate, but above that minimum there is not a strong frequency dependence. For low $\omega$, shock heating occurs after $\alpha$ recombination is complete and seed formation is well underway, and thus has a negligible effect on nucleosynthesis. For a sufficiently high $\omega$, the waves shock early enough to disrupt $\alpha$ recombination, allowing for an r-process. }
\end{figure*}
\begin{figure}
    \centering
    \includegraphics[scale=1]{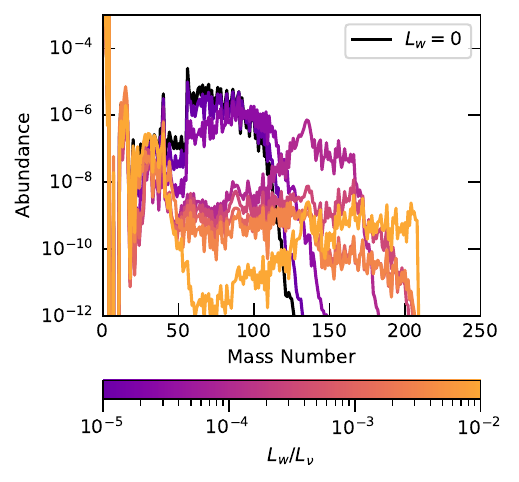}
    \caption{Final nucleosynthesis results, using temperature and density profiles for a 1.5 $M_\odot$ neutron star, with $L_\nu=3\times 10^{52}$ erg s$^{-1}$, $\omega=2\times 10^3$ rad s$^{-1}$, and assuming shock heating begins immediately in the wind (i.e. $R_s=R_\text{NS}$). We observe a smoother transition from a $\nu$p-process at low $L_w$ to a fast-outflow r-process at high $L_w$.}
    \label{fig:fid_instant}
\end{figure}

\begin{figure*}
\begin{subfigure}{.5\textwidth}
    \centering
    \includegraphics[scale=1]{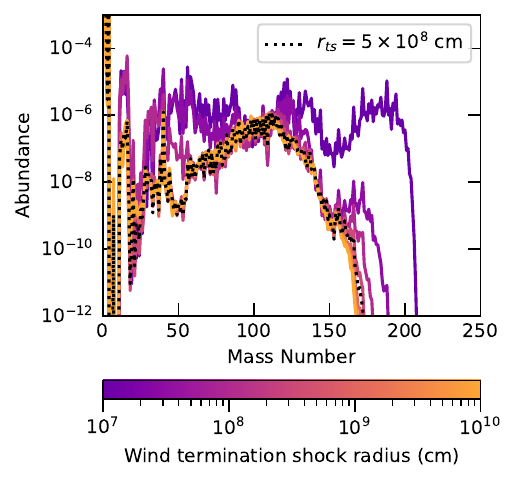}
    \caption{$\nu$p-process ($L_w/L_\nu=1\times 10^{-4}$)}
    \label{fig:rev_vp}
\end{subfigure}%
\begin{subfigure}{.5\textwidth}
    \centering
    \includegraphics[scale=1]{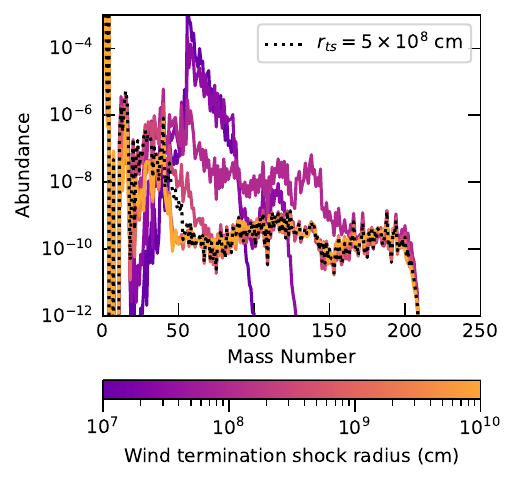}
    \caption{r-process ($L_w/L_\nu=5\times 10^{-3}$)}
    \label{fig:rev_r}
\end{subfigure}
\caption{Final nucleosynthesis results for a fiducial 1.5~$M_\odot$ PNS with $L_\nu=3\times 10^{52}$ erg s$^{-1}$ and $\omega=2\times 10^3$ rad s$^{-1}$, highlighting the wind termination shock location ($r_\text{ts}$) dependence of the two nucleosynthetic processes. The $r_\text{ts}=5\times 10^8$~cm case used in our other plots is highlighted for comparison. Left: $L_w/L_\nu=1\times 10^{-4}$ to produce a $\nu$p-process. An earlier shock is beneficial for the $\nu$p-process, keeping the material in the optimal 1.5-3 GK temperature window that \citet{Wanajo_2011} and \citet{Arcones_2012b} observed. Right: $L_w/L_\nu=5\times 10^{-3}$ to produce a fast-outflow r-process. An early shock can delay $\alpha$ capture freezeout by keeping the material at high temperatures longer, leading to rampant seed formation and stifling the r-process. A later shock allows the wind to cool, "freezing in" the low seed count caused by the additional wave heating and allowing the fast-outflow r-process to proceed. \citet{Kuroda_2008} similarly observed that a later termination shock is optimal for r-processing.}
\end{figure*}

We now consider the nucleosynthesis that results from the wind models described in the preceding section.
Figure \ref{fig:fid_skynet} shows the final abundances produced by the fiducial wind profiles shown in figure \ref{fig:fid_profiles}. Two distinct families emerge in the abundance patterns. For low to moderate $L_w$, we see a structure that is generally similar to the $\nu$p-process patterns that emerge from the $L_w=0$ case, with a broad peak near $A=100$ that shifts towards higher mass numbers with increasing $L_w$. As in the neutron-rich case, heavy nucleosynthesis is suppressed in a small window near $L_w/L_\nu=10^{-3}$ due to competition between reduced entropy and reduced dynamical timescale causing increased seed formation (see Paper I). At $L_w \gtrsim 10^{-3} L_\nu$, the waves shock prior to seed formation, allowing again for heavier nucleosynthesis. We see what could be described as a suppressed r-process emerge, reaching to mass numbers near $A=210$. 

For a more extreme parameter set, with higher mass and $L_\nu$, we observe similar abundance families. Figure \ref{fig:ext_skynet} shows the abundances produced by this parameter set with a higher PNS mass ($M_\text{NS}=1.9 M_\odot$) and neutrino luminosity ($L_\nu=6\times 10^{52}$ erg s$^{-1}$). Under these conditions, the $\nu$p-process is able to proceed all the way to $A\approx 200$, with the peak shifting to near $A=175$ for moderate $L_w$. Heavy nucleosynthesis is again suppressed near $L_w/L_\nu=10^{-3}$, and revives when shock heating begins to affect seed formation. The suppressed r-process pattern again emerges at high $L_w$, albeit with lower overall abundances.

The diagnostic quantity $s^2/\tau$ \citep[similar to $s^3 Y_e^{-3}\tau^{-1}$ in neutron-rich winds, see][]{Hoffman_1997} is helpful in explaining how the strength of the $\nu$p-process varies with $L_w$, $L_\nu$, and mass. The entropy dependence of seed formation is reduced by one power relative to the neutron-rich case since seed formation proceeds through the triple-alpha pathway in proton-rich conditions rather than the effective four-body reaction $^{4}\textrm{He}(\alpha, \gamma)^{8}\textrm{Be}(n,\gamma)^{9}\textrm{Be}(\alpha, n)^{12}\textrm{C}$. The $\nu$p-process, just like the r-process, is limited by the free-nucleon-to-seed-nucleus ratio. A higher $s^2/\tau$ implies a higher proton-to-seed ratio, which will allow heavier nuclei to be formed. Figure \ref{fig:s3tau} shows the correlation, especially for low to moderate $L_w$, between $s^2/\tau$ and the heaviest nuclide formed in the wind. Although the two quantities are correlated throughout most of the $L_w/L_\nu$ range, in the range $2\times 10^{-4}\lesssim L_w/L_\nu\lesssim 10^{-3}$, a discrepancy between the two arises. This is driven by two factors. First, the faster outflow due to wave heating at smaller radii reduces the electron fraction of the outflow at the beginning of nucleosynthesis (see Paper I). Second, after seed formation a shorter dynamical timescale implies there is a shorter period of time for $\textrm{p}(\nu_e, e^+)\textrm{n}$ to produce neutrons to bypass long-lived waiting point nuclei before the $r^{-2}$ falloff of the neutrino flux shuts off further captures. Our simulations indicate that this second effect is dominant, and the reduced electron fraction plays a lesser role. To illustrate this, figure \ref{fig:nv} shows the reduction in antineutrino captures (i.e. the electron antineutrino capture rate integrated from $T=3$~GK through the end of the simulation) in the $\nu$p-processing regions of the wind as the wave luminosity is increased.  Above this range of $L_w$, shock heating begins early enough to be the dominant effect in the wind and drives suppressed r-processing up to $A\gtrsim 200$.

A closer analysis of the nucleosynthesis calculations reveals that the abundance patterns we observe are the product of several different processes that take place at different temperatures. For lower wave luminosities, a robust $\alpha$-recombination takes place, and we observe a $\nu$p-process taking place at temperatures near 3 GK, followed by a neutron capture epoch when the wind enters ($n,\gamma$) - ($\gamma,n$) equilibrium, driving the final abundances to the neutron-rich side of stability. This matches the predictions of \citet{Pruet_2006} for winds with enhanced entropy. The $\nu$p-process is very entropy-dependent, so we do not see nucleosynthesis up to $A=200$ in the relatively low-entropy fiducial case. At higher PNS masses (i.e. higher entropies) it can proceed farther, and produces the peak near $A=175$ that we see in figure \ref{fig:ext_skynet}. In the range $2\times 10^{-4}\lesssim L_w/L_\nu\lesssim 10^{-3}$, we see the $\nu$p-process is curtailed in both the fiducial and extreme cases. This is due to both the increased seed formation we previously observed in this region (Paper I), as well as a reduction in neutron production. The additional acceleration from wave stresses reduces the number of neutrino captures at early times, keeping $Y_e$ closer to neutron richness and reducing the number of free protons available for neutron production during $\nu$p-processing. Additionally, the faster outflow reduces the time in which neutron production via antineutrino capture can proceed, decreasing total neutron production during $\nu$p-processing. This reduction is shown in figure \ref{fig:nv}.

At the highest wave luminosities, the early shock heating prevents a robust nucleon recombination phase, leaving a noticeable abundance of both free protons and free neutrons and drastically reducing seed production. Running the nucleosynthesis calculations without neutrino reactions shows that these neutrons are not created in-situ, confirming that this is not a $\nu$p-process; the bulk of the neutrons are left over from an incomplete nucleon recombination phase. As the temperature continues to drop, charged particle reactions freeze out and a neutron capture epoch ensues. In the fiducial parameter set (figure \ref{fig:fid_skynet}), the lower neutrino luminosity leads to a lower electron fraction, providing a larger reservoir of neutrons, which results in more pronounced r-process peaks near $A=140$ and 200. In the more extreme parameter set (figure \ref{fig:ext_skynet}), the higher neutrino luminosity reduces the number of free neutrons, and thus the r-process pattern is more suppressed. The abundance patterns for the high $L_w$ case, especially those in figure \ref{fig:ext_skynet}, agree very closely with those predicted by \citet{Meyer_2002} for r-processing in very fast outflows. It is important to note that, as figure \ref{fig:fid_profiles} indicates, the r-processing here occurs in proton-rich conditions, with antineutrino energies that predict an equilibrium $Y_e$ of 0.6.

Incidentally, reviewing the high $L_w$ nucleosynthesis calculations for neutron-rich winds (e.g. figure 7 of Paper I) we find that this same process was operating in those winds. We did not see the same final abundance patterns because the free neutron abundance was high enough that the subsequent period of neutron capture washed out any trace of the earlier process, yielding a standard r-process abundance distribution.  

\subsubsection{Wave Frequency}
The frequency of the waves excited by PNS convection plays an important role in determining where and how the waves will shock. Though there is substantial uncertainty in the shock mechanism in this context, we expect higher frequencies to shock earlier and deposit their energy faster than lower frequencies (see equations 13 and 16 of Paper I, and the scaling relations in section \ref{sec:theory} of this work). In figures \ref{fig:freq_vp} and \ref{fig:freq_r}, we present a typical $\nu$p- and fast outflow r-process with varied wave frequencies, as well as one abundance profile from a wind in which we assume the waves shock instantly in the wind, to illustrate the effect of the uncertainty in the shock formation prescription. In the case of the $\nu$p-process, we see very little frequency dependence because of the low wave luminosity, which limits how early the shocks can form in the main prescription we assume (see Paper I, section 3.2). Even when the shocks form instantly, the waves are not carrying enough energy to strongly alter the nucleosynthesis. Since the $\nu$p-process only operates at lower wave luminosities ($L_w/L_\nu\lesssim 10^{-3}$), we expect this to be consistent across the parameter space.

The fast-outflow r-process shows a stronger frequency dependence, in the form of what appears to be a minimum cutoff frequency. Because this r-process only operates at high wave luminosities ($L_w/L_\nu\gtrsim 10^{-3}$), if the waves shock early enough, they will deposit enough energy to strongly affect the resulting nucleosynthesis. At very low frequencies, the waves do not shock until seed formation has already taken place, so the r-process is stifled. Once above the cutoff frequency, the waves shock early enough to disrupt $\alpha$ recombination, and the final abundance pattern is not strongly affected by further increasing the frequency. 

There is substantial uncertainty in the shock mechanism for the waves we are considering, and they may shock much earlier than predicted (see figure 10 and the discussion in Paper I). In figure \ref{fig:fid_instant}, we present abundance profiles assuming the waves shock instantly, so that any heat deposition will affect every stage of nucleosynthesis. We observe what appears to be a smoother transition from a $\nu$p-process at low wave luminosities, to a fast-outflow r-process at high wave luminosities, with higher free neutron abundances altering the abundance patterns as $L_w$ is increased.

\subsubsection{The Role of the Wind Termination Shock}
Because of the strong dependence of the $\nu$p-process on the location of the wind termination shock found in previous work \citep[e.g.][]{Arcones_2012b}, we now turn our attention there. Figure \ref{fig:rev_vp} shows how the final abundance pattern for a typical $\nu$p-process varies with the location of the wind termination shock. Previous studies have found that an early wind termination shock at temperatures of around 2 GK is optimal for facilitating a strong $\nu$p-process. Our wind models cool very rapidly, so this optimal temperature range occurs very early on. We can see that, as expected, the smallest shock radius is favored for the $\nu$p-process. An early shock keeps the wind at higher temperatures and smaller radii longer, which allows more time for both proton captures and neutron production via antineutrino capture to proceed (see figure \ref{fig:nv}). 

For the fast-outflow r-process, a later shock is optimal. This is in at least qualitative agreement with \citet{Kuroda_2008}, who observed less robust r-processes with early shock radii, and little change in abundance patterns once the radius was increased above 3000~km. We see in figure \ref{fig:rev_r} that the r-process cannot proceed for very early termination shocks. A sufficiently early termination shock can occur at high enough temperatures to restart $\alpha$ capture reactions after their initial freezeout, and the slow-moving shocked material takes much longer to cool than the fast-moving wind. Seed formation proceeds rapidly in such cases, and we see a robust $\alpha$ process emerge, forming large amounts of iron-peak elements and precluding an r-process. Once the shock is moved beyond approximately 1000~km the r-process can proceed with full vigor, and when the shock is moved beyond approximately 3000~km the final abundance pattern is no longer sensitive to its location. 

\subsection{Nucleosynthesis in the $L_w$ - $L_\nu$ - $M_\text{NS}$ parameter space}

\begin{figure*}
    \centering
    \includegraphics[scale=1]{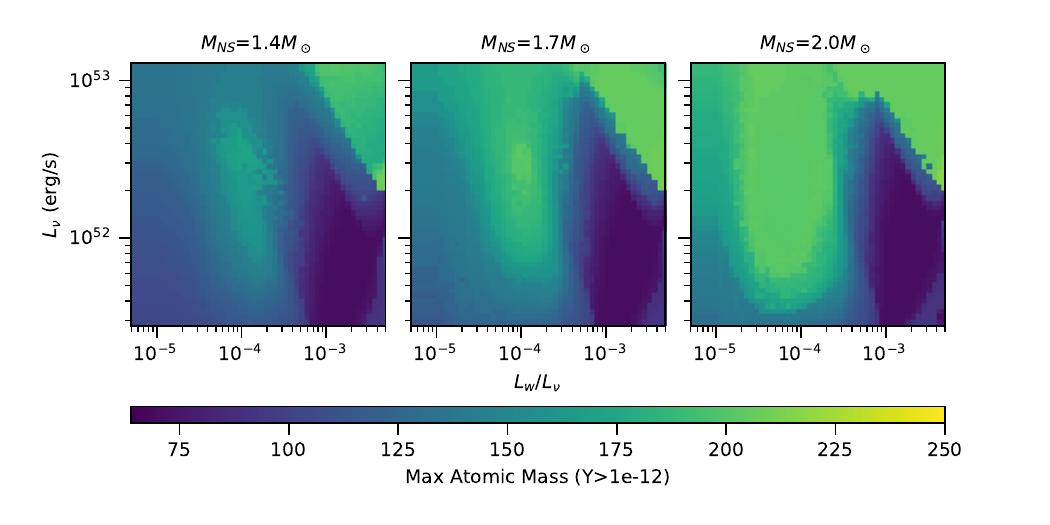}
    \caption{This plot shows the highest atomic mass produced in a significant amount ($Y\geq 1\times 10^{-12}$), giving a measure of how far heavy element nucleosynthesis is able to proceed. We observe similar heavy nucleosynthesis regimes as in the neutron-rich case, with a mass-dependent wave-stress driven $\nu$p-process regime at moderate wave luminosities, and a shock-heating driven, mass-independent r-processing regime at high $L_w$ and $L_\nu$. We do not observe any actinide production in these models.}
    \label{fig:default_max}
\end{figure*}
\begin{figure*}
    \centering
    \includegraphics[scale=1]{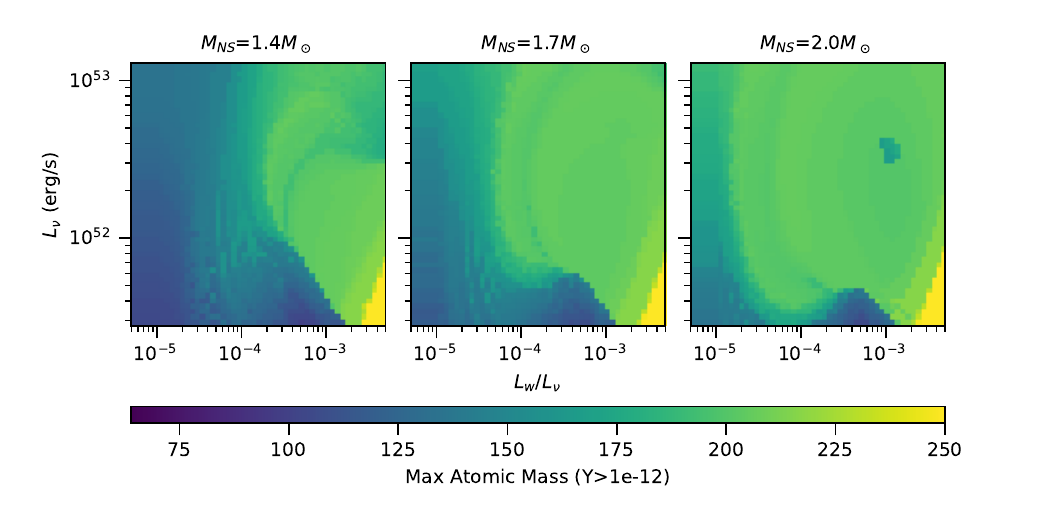}
    \caption{This plot is the same as figure \ref{fig:default_max}, but assuming that shock heating begins immediately in the wind. Similar to the neutron-rich case, we see a smoother transition between and blending of the wave-stress and shock-heating regimes of heavy nucleosynthesis. The high $L_w$, low $L_\nu$ region of strong heavy element production is revealed to be a full r-process proceeding all the way to the actinides, as wave effects drive the wind to neutron-richness.}
    \label{fig:instant_max}
\end{figure*}

In figure \ref{fig:default_max}, we show a plot of the maximum atomic mass that is produced with a final abundance greater than $10^{-12}$ for a given parameter set. This gives a helpful measure of how far the $\nu$p-process is able to proceed as a function of PNS mass, neutrino luminosity, and wave luminosity as a fraction of neutrino luminosity. For these plots, we have taken $\omega=2000$~s$^{-1}$, $Y_{e,\text{eq}}=0.6$, $r_\text{ts}=5\times 10^8$~cm, and used the shock prescription in equation \ref{eq:rshock}. We observe similar nucleosynthesis regimes as in the neutron-rich case in paper I. For high $L_\nu$ and $L_w$, we see very heavy nuclei being formed as shock heating begins early enough to inhibit seed formation, and the high entropy and fast outflow prevent a complete $\alpha$-recombination, leading to a fast-outflow r-process. Because the fast-outflow r-process depends on early shock heating by the waves disrupting $\alpha$-recombination, the sharp cutoff line of this region indicates that the minimum wave luminosity to obtain such an r-process is of order $10^{50}$~erg~s$^{-1}$, and is relatively independent of neutrino luminosity. We also observe no actinide production in any region of this parameter space, which could lead to an observable signature of these winds. 

We also see a PNS mass-dependent nucleosynthesis regime driven by the wave stress emerge, as in the neutron-rich case. The nucleosynthesis patterns here are broadly consistent with a $\nu$p-process, and exhibit the expected correlation between the wave luminosity, the diagnostic quantity $s^2/\tau$, and the maximum atomic mass produced shown in figure \ref{fig:s3tau}. The non-monotonic connection between wave luminosity and maximum mass produced is very clearly illustrated here. At low wave luminosities, the effect on $s^2/\tau$ dominates, and the reduction in seed formation allows more massive nuclides to form. For $L_w/L_\nu\gtrsim 5\times 10^{-4}$, unless the neutrino luminosity is high enough to make $L_w\gtrsim10^{50}$~erg~s$^{-1}$ and drive fast-outflow r-processing, the reduction in neutrino captures caused by the additional acceleration becomes dominant, and the $\nu$p-process is stifled by a lack of free neutrons. Nucleosynthesis shifts toward the iron peak, and heavy nuclides cannot form. Both the fast-outflow r-process and the $\nu$p-process benefit from higher asymptotic entropies, so we see a positive correlation between PNS mass and the maximum atomic mass nuclides formed in the wind.

Because of the uncertainty surrounding when and where shock formation will take place in the wind, in figure \ref{fig:instant_max} we show the same parameter space as in figure \ref{fig:default_max}, but assuming that the waves instantly shock and deposit their energy into the wind. Just as in the neutron-rich case, we see the shock-heated heavy nucleosynthesis regime (here the fast-outflow r-process) broadens in both $L_w$ and $L_\nu$, and the wave pressure regime becomes difficult to distinguish as it emerges. Because shock heating is assumed to begin instantly, $\alpha$-recombination is affected even at low wave luminosities, and we see a smooth transition from $\nu$p-processing to fast-outflow r-processing as in figure \ref{fig:fid_instant}. At high masses (i.e. higher entropies), the parameter space is dominated by heavy nucleosynthesis of some kind. Interestingly, we notice a new regime emerge when instant shock heating is assumed: for low $L_\nu$ and high $L_w$, the wind material becomes neutron-rich, and a full, third-peak, actinide-producing r-process appears. The rapid acceleration and additional heat provided by the waves, coupled with a low neutrino luminosity, result in a wind that undergoes comparatively few charged-current neutrino interactions and maintains a $Y_e$ well below its equilibrium value. We emphasize again that the neutrino spectra of these proto-neutron stars predicts an equilibrium electron fraction of 0.6, yet we observe regions of the parameter space (albeit extreme) that predict a full r-process.

\section{Conclusions}
In this paper, we have continued our exploration of the effects of convection-driven acoustic waves on nucleosynthesis in the neutrino-driven wind, now focusing on winds predicted to be proton-rich and with an eye toward the $\nu$p-process. As in Paper I, we have surveyed the parameter space of PNS masses, neutrino luminosities, wave luminosities, and wave frequencies, as well as considering the impact of the wind termination shock and different shock heating prescriptions. Using a steady-state, one-dimensional wind code and the SkyNet reaction network, we have performed nucleosynthesis calculations to determine the predicted abundance profiles and reaction flows expected in these winds. For the bulk of our calculations, we assume a PNS neutrino spectrum that yields an equilibrium $Y_e$ of 0.6, in the range predicted by many recent simulations.

As in the neutron-rich case, we find that wave luminosities of $L_w\gtrsim 10^{-5}L_\nu$ have a substantial impact on wind dynamics and nucleosynthesis. At modest wave luminosities ($L_w\lesssim 10^{-3}L_\nu$), the faster expansion caused by the wave stress contribution decreases seed formation efficiency, leaving a higher ratio of free protons to seed nuclei and enhancing the $\nu$p-process. The resulting nucleosynthesis flows are able to produce elements up to mass 208 and beyond, though not reaching the actinides. This is a strong enhancement from our models that neglect wave effects, which only reach up to around mass 130. 

We find these wave driven effects at modest luminosities can have competing impacts on $\nu$p-process nucleosynthesis in the NDW: 
\begin{enumerate}
\item Gravito-acoustic wave driven acceleration and shock driven energy deposition impacts nuclear seed production for the $\nu$p-process. Acceleration of the NDW due to wave stresses both reduces the dynamical timescale and also reduces the amount of neutrino heating that occurs (which in turn reduces the asymptotic entropy of the NDW). Additionally, after the gravito-acoustic waves become non-linear and shock, they deposit heat in the NDW and increase the entropy. Since seed production in proton-rich winds depends on the parameter $s^2/\tau$, with seed production decreasing with increasing values of this parameter, the impact of waves on seed production can behave non-monotonically with $L_w$. For $L_w \lesssim 2 \times 10^{-4} L_\nu$, we find the decreased dynamical timescale dominates and seed production decreases with increasing $L_w$. For $2 \times 10^{-4} L_\nu \lesssim L_w \lesssim 2 \times 10^{-3}$, seed production increases with $L_w$ as the entropy reduction due to wave stresses dominates. For $L_w \gtrsim 2 \times 10^{-3} L_\nu$, wave shock heating increases both the entropy and dynamical timescale to the extent that even alpha production is impacted.  

\item Acceleration of the NDW by gravito-acoustic waves reduces the time available for $\textrm{p}(\bar \nu_e, e^+)\textrm{n}$ to occur and thus reduces the total number of neutrons that are produced per seed nucleus. This reduces the maximum nuclear mass to which the $\nu$p-process proceeds, but at lower wave luminosities the effects on seed formation via the increased entropy and decreased dynamical timescale dominate. At $2\times 10^{-4}\lesssim L_w/L_\nu\lesssim 10^{-3}$, the reduced neutron flux becomes significant enough to curtail $\nu$p-processing somewhat.

\item For larger gravito-acoustic wave luminosities, the acceleration of the NDW by these waves can reduce the number of neutrino captures that occur before alpha particle formation. Since material in the PNS atmosphere is in beta equilibrium and as a result is very neutron rich, the reduced number of neutrino captures as the outflow starts tends to decrease the proton richness of the wind. This effect was seen for neutron rich winds, where it helped to make conditions more favorable for r-process nucleosynthesis, but in proton rich winds where the $\nu$p-process might occur this effect reduces the number of free protons that are available to undergo $\textrm{p}(\bar \nu_e, e^+)\textrm{n}$ once seeds have been formed. This also contributes to the reduction in maximum nuclear mass attained by the $\nu$p-process seen for moderate wave luminosities. 
\end{enumerate}

At higher wave luminosities ($L_w\gtrsim 10^{-3}L_\nu$) and higher neutrino luminosities ($L_\nu\gtrsim 10^{52}$ erg s$^{-1}$), we find an interesting result: the early shock heating in this region of the parameter space prevents a complete nucleon recombination phase in the wind, preserving an abundance of free neutrons much higher than the persistent abundance created by charged-current neutrino interactions. This, coupled with the reduced seed formation due to higher entropy, allows a suppressed r-process to take place, similar to that predicted by \citet{Meyer_2002} for fast-moving winds. Though this r-process does not reach to the actinides, it can reach to the third r-process peak and beyond. For a substantial region of the parameter space, we find that winds from a neutron star with a neutrino spectrum predicted to result in a firmly proton-rich wind can undergo an (albeit suppressed) r-process. When we consider an alternative prescription for the shock heating, in which the waves immediately shock upon entering the wind, not only does this region of suppressed r-processing broaden, but we find an even more extreme possibility: the combination of acceleration due to wave stresses and additional heating from the shocked waves reduces the amount of neutrino heating needed to unbind the wind to such a degree that the wind actually becomes neutron-rich, and undergoes a full, third-peak r-process reaching the actinides. While this only occurs in a small corner of the parameter space, its mere possibility demonstrates the dramatic changes that can be caused by the presence of waves in the NDW. 

Many uncertainties in the long-term evolution of PNS neutrino spectra and convective properties, and in the mechanics of wave propagation in this environment render our models necessarily approximate, and our parameter-survey approach here precludes detailed, time-integrated production estimates of individual nuclei of interest such as the light p-nuclides. However, our results clearly indicate the significant effects of gravito-acoustic waves being present in the NDW. As late-time supernova models continue to improve, accurate nucleosynthesis calculations will require well-resolving the convective properties of the proto-neutron star and any waves propagating through the neutrino-driven wind. The recent 3D simulations of \citet{Wang_2023} extending to late times are an important step in this process, and show that the transonic winds we consider here do obtain in more detailed simulations. We look forward to calculating time-integrated nucleosynthesis for a realistic PNS evolution to compare with their results in a future paper, with an eye towards the production of the light p-nuclides $^{92,94}$Mo and $^{96,98}$Ru specifically.

\section*{Acknowledgements}
BN thanks Edward Brown for helpful discussions during this work. BN acknowledges support from a University Distinguished Fellowship and from the College of Natural Sciences at Michigan State University. This work was supported in part through computational resources and services provided by the Institute for Cyber-Enabled Research at Michigan State University.

\section*{Data Availability}
The simulation code and results used in this work are available upon reasonable request to the authors. The SkyNet reaction network used is open-source software publicly available at \url{https://bitbucket.org/jlippuner/skynet}.





\bibliographystyle{mnras}
\bibliography{neutrino_winds_bibliography} 





\bsp	
\label{lastpage}
\end{document}